\documentclass[reprint,pra,longbibliography,superscriptaddress,nofootinbib]{revtex4-2}
\usepackage[utf8]{inputenc}
\usepackage{lipsum}
\usepackage{graphicx}
\usepackage[caption=false]{subfig}
\usepackage{braket}
\usepackage[dvipsnames]{xcolor}
\usepackage{mathtools}
\usepackage{amsmath}
\usepackage{amssymb}
\usepackage{hyperref}
\usepackage{esvect}
\usepackage{float}
\usepackage[normalem]{ulem}
\graphicspath{{./PaperFigures/}}

\allowdisplaybreaks

\hypersetup{
    colorlinks=true,
    linkcolor=blue,
    filecolor=magenta,      
    urlcolor=cyan,
    citecolor={blue},
    }

\begin{document}

\title{Quantum-dot single photon source performance with off-resonant pulse preparation schemes}

\author{Gavin Crowder}
\email{gcrow088@uottawa.ca}
\affiliation{Department of Physics, University of Ottawa, Ottawa, Ontario, Canada, K1N 6N5}
\affiliation{Nexus for Quantum Technologies Institute, University of Ottawa, Ottawa, Ontario, Canada, K1N 6N5}
\affiliation{Department of Physics, Queen's University, Kingston, Ontario, Canada, K7L 3N6}
\author{Lora Ramunno}
\affiliation{Department of Physics, University of Ottawa, Ottawa, Ontario, Canada, K1N 6N5}
\affiliation{Nexus for Quantum Technologies Institute, University of Ottawa, Ottawa, Ontario, Canada, K1N 6N5}
\author{Stephen Hughes}
\affiliation{Department of Physics, Queen's University, Kingston, Ontario, Canada, K7L 3N6}

\begin{abstract}
The preparation of photonic qubits in the excited state is an integral part of the performance of an on-demand single photon source (SPS). Conventional resonant excitation, an excellent approach to maximize the coherence and indistinguishability of the SPS, often requires polarization filtering to remove the pump signal and isolate the qubit emission, but this results in an inherent 50\% hit to the efficiency. Recent excitation schemes strategically try to exploit pulses that excite the qubit while avoiding spectral overlap to bypass this required filtering. In this work, we compare three such pumping schemes to quantify the important SPS figures-of-merit for off-resonant quantum dot schemes, using: (i) a symmetrically detuned dichromatic pulse, (ii) a notch-filtered adiabatic rapid passage (NARP) pulse, and (iii) a swing up of the quantum emitter population (SUPER) pulse. Due to large instantaneous pulse strengths, the dichromatic pulse suffers from phonon-induced dephasing which can lower the SPS performance by up to 50\%. In contrast, the NARP and SUPER pulses are shielded from phonon coupling to differing degrees but both maintain excellent SPS performance. The SUPER pulse can lose significant efficiency if there is variance in its constituent pulses' amplitude, pulse width, or frequency, while the NARP pulse, though potentially more difficult to realize in experiments, is robust against variance in the pulse preparation.
\end{abstract}
\maketitle

\section{Introduction}
\label{sec:Intro}

A high performance on-demand single photon source (SPS) is an invaluable resource for fundamental quantum optics experiments and quantum technologies \cite{Kiraz2004}. An ideal deterministic source (i.e., on-demand) would be perfectly efficient ($\eta = 1$), meaning it emits every time it is pumped and the emitted photons would be coherent ($g^{(2)}(0) = 0$) and indistinguishable ($\mathcal{I} = 1$). The efficient generation of pure quantum light is an ongoing technical problem being worked on in many different platforms, for example superconducting transmon qubits~\cite{Pechal2014,Zhou2020,Lu2021}, trapped ions~\cite{Higginbottom2016,Crocker2019}, spin-centres in solid state sources such as diamond~\cite{Englund2010,Grosso2017,Schrinner2020}, and quantum dot (QD) sources~\cite{Takemoto2004,He2013,Kalliakos2014,Wei2014,Schweickert2018,Schnauber2019,Northeast2021,Ding2025} to list a few, each with their own approaches to performance improvement. 

Current state of the art QD SPSs have achieved coherences as low as $g^{(2)}(0) = 7.5\times 10^{-5}$~\cite{Schweickert2018} and indistinguishabilities as high as $\mathcal{I} = 0.995$~\cite{Wei2014}. Simultaneously maximizing these figures of merit, while keeping a high efficiency is a difficult problem with a recent paper showing an overall best performance (to our knowledge) of $g^{(2)}(0) = 0.0205$, $\mathcal{I} = 0.9856$, and $\eta = 0.712$~\cite{Ding2025}. These near pure SPSs are typically driven by resonant $\pi$-pulses that coherently drive inversion of the qubit~\cite{Stievater2001,Ding2016}. The 
potential problem with this scheme in a waveguide-QED system is that the weak qubit emission and driving pulse both travel through the same waveguide channel causing the emission to be overshadowed by the much stronger driving pulse. To overcome this, techniques such as polarization filtering are used~\cite{NickVamivakas2009,Claudon2010,Ding2016,Somaschi2016,Liu2019} where the light with polarization parallel to the driving pulse polarization is filtered to isolate the qubit emission. This inherently removes half of the qubit emission events, capping $\eta$ at 0.5 which does not meet the requirement of an on-demand source.



To avoid this required filtering, it would be useful to have a driving scheme which avoids using a laser on resonant with the qubit. There have been many such driving protocols such as off-resonant phonon-assisted inversion~\cite{Ardelt2014,Quilter2015,Manson2016,Reindl2019,Cosacchi2019,Thomas2021}, two photon excitation of the biexciton~\cite{Brunner1994,Muller2014,Schweickert2018,Hanschke2018}, and p-shell excitation~\cite{Ware2005,Gazzano2013,Laferriere2022}. Such schemes have shown excellent results, with a CW-dressing scheme utilizing the biexciton-exciton cascade yielding $g^{(2)}(0) = 0.028$, $\mathcal{I} = 0.905$, and $\eta > 0.95$~\cite{Dusanowski2022,Gustin2022}. This technique avoids polarization filtering but introduces timing jitter (decoherence due to the finite spontaneous emission time of the biexciton-exciton decay) which must be overcome~\cite{Gustin2020,Scholl2020}. Recently, three such schemes have been proposed and experimentally demonstrated, showing only a small effect on the SPS performance by preparing the qubit with these protocols. These are a symmetrically detuned dichromatic pulse protocol~\cite{He2019,Koong2021}, the notch filtered adiabatic rapid passage (NARP) protocol~\cite{Wilbur2022}, and the swing up of the quantum emitter population (SUPER) protocol~\cite{Bracht2021,Karli2022,Bracht2023,Torun2023,Heinisch2024}.

These three schemes have been presented with a variety of simulated and experimental results on different platforms to show their performance as a SPS.
However, these various studies all make different underlying assumptions for their systems (such as different dissipation modeling and qubit decay rates) which makes a direct comparison between these works difficult. It is useful for the scientific community to compare the performance of these schemes in a realistic system while removing as many differences in the implementation as possible. Indeed in~\cite{Vannucci2023}, a comparison between the dichromatic and SUPER schemes is done for a qubit coupled to a cavity which drives emission into the zero phonon line. In addition, we are not aware of any works that have computed $\mathcal{I}$
for the NARP schemes. 

\begin{figure}
    \centering
    \includegraphics[width=0.85\columnwidth]{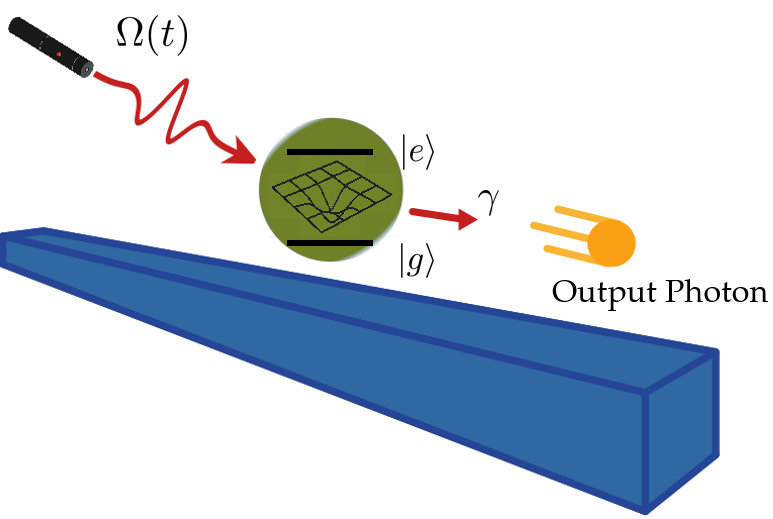}%
    \caption{Schematic of the single photon emission process. The qubit (with LA acoustic phonons and coupled to the waveguide with rate $\gamma$) is driven by one of the three pulse schemes and (ideally) emits a single photon in the waveguide.}
    \label{Schematic}
\end{figure}

In this paper, we present a theoretical study of a SPS platform, where we consider a QD (treated as the qubit) generating single photons in a waveguide quantum electrodynamic (QED) system shown schematically in Fig.~\ref{Schematic}. Our model also accounts for phonon-induced processes, in the calculations of the key SPS parameters.

Specifically, we make a detailed comparison of the SPS performance of a bare qubit coupled to a waveguide under the dichromatic, NARP, and SUPER schemes. To incorporate dissipation from a realistic system, we also employ a weak phonon coupling model (appropriate for low temperatures) \cite{Gustin2020} and plot the resulting phonon-induced dissipation rates which arise from the drive strength and detuning of each of the three schemes. Using parameters from the original publications for each driving scheme, we compare the SPS performance in $\eta$, $g^{(2)}(0)$, and $\mathcal{I}$. These are the first calculations of indistinguishability for the NARP protocol that compare its performance against other driving schemes. Additionally, the robustness of each scheme is considered, that is if the incoming laser pulse (or constituent pulses) are imperfect, how much does that impact the performance of the SPS under each driving scheme.

Our key findings show that the dichromatic scheme suffers from the influence of phonons, which significantly lowers its performance as a single photon source. The NARP scheme has limited effects from phonon coupling while the SUPER scheme completely avoids the problem with its far detuned pulses. We also find that both NARP and SUPER both have excellent coherence and indistinguishability in the resulting single photons. While the SUPER scheme has slightly better overall performance, it is highly sensitive to changes in the composite laser parameters, while the NARP scheme is quite robust against variation in the filtering width or laser amplitude.

\section{Theory and Methods}
\label{sec:Theory}

\subsection{System Hamiltonian}
\label{sec:Model}

To investigate the performance of each of the driving schemes, the SPS is modeled as a QD exciton (single electron-hole pair) including its coupling to longitudinal acoustic (LA) phonons. Specifically, the QD is treated as a two level system (TLS), which we refer to as a qubit, with ground state $\ket{g}$, excited state $\ket{e}$, and resonance frequency $\omega_0$. Neglecting electron-phonon coupling and dissipation processes for now, the system Hamiltonian for an optically excited quantum dot is 
\begin{equation}
    H_{\rm{S}} = \hbar\omega_0 \sigma^+ \sigma^- + \frac{\hbar\Omega(t)}{2}(\sigma^+ + \sigma^-),
    \label{Sys_Ham}
\end{equation}
where $\sigma^+$ ($\sigma^-$) is the Pauli raising (lowering) operator. The time-dependent Rabi field $\Omega(t)$ is defined by the driving field of interest, before we have moved into an interaction picture and without any rotating wave approximations.

The qubit is coupled to a waveguide with radiative rate $\gamma$, proportional to the dipole strength and the local density of states in the waveguide. This is included with the quantum jump channel $C_0 = \sqrt{\gamma}\sigma^-$ in a quantum trajectory picture~\cite{Dum1992,Tian1992,Dalibard1992}, equivalent to a Lindblad dissipation operator in a master equation approach.

\subsection{Optical Pulsed Pumping Schemes}
\label{sec:Schemes}

We now present the specific form for the system Hamiltonian in Eq.~\eqref{Sys_Ham}, in the rotation wave approximation, for the three pumping schemes we consider here: the dichromatic, NARP, and SUPER pulses, along with a mathematical description of each scheme.

The first of the three pulsed driving schemes we compare, is a symmetrically detuned dichromatic pulse proposed in~\cite{He2019} that consists of two pulses, each symmetrically detuned from the TLS by $\delta$. The two pulses have identical envelopes, $\varepsilon(t)$, are coincident in time, but can have a fixed phase difference $\varphi$. The total Rabi frequency for these two pulses is
\begin{equation}
    \begin{aligned}
    \Omega(t) & {}= \frac{\hbar \varepsilon(t)}{2} \cos \left( (\omega_{\rm{L}} + \delta) t + \varphi \right) + \frac{\hbar \varepsilon(t)}{2} \cos \left( (\omega_{\rm{L}} - \delta) t \right) \\
    & {}= \hbar\varepsilon(t) \cos \left( \delta t + \frac{\varphi}{2} \right) \cos \left( \omega_{\rm{L}} t + \frac{\varphi}{2} \right).
    \end{aligned}
\end{equation}

In the second line we can see that these two pulses act like a resonant pulse with a new envelope $\varepsilon'(t) = \varepsilon(t) \cos \left( \delta t + \frac{\varphi}{2} \right)$. There is a small difference from the resonant case introduced by $\varphi$ in the $\cos \left( \omega_{\rm{L}} t + \frac{\varphi}{2} \right)$ term. When moving into the interaction picture at the laser frequency with a rotating wave approximation (RWA), this introduces a phase on the TLS operators 
\begin{equation}
    H_{\mathrm{S}}^{\rm Di} = \frac{\hbar\varepsilon'(t)}{2} \left( e^{-i\varphi/2} \sigma^+ + e^{i\varphi/2} \sigma^- \right).
\end{equation}

To yield population inversion, as predicted by the Area Theorem (and neglecting dissipation processes), the time integrated pump area must be an odd multiple of $\pi$, i.e. $\int dt \varepsilon'(t) = (2n + 1)\pi$ where $n \in \mathcal{Z}$. If we take the original pulse envelopes to be Gaussian,
\begin{equation}
    \varepsilon(t) = \Omega_0 e^{-t^2/2t_{\rm p}^2},
\end{equation}
then we need to choose
\begin{equation}
    \Omega_0 = \frac{\pi}{\sqrt{2\pi t_{\rm p}^2} \cos(\varphi/2) e^{-\delta^2 t_{\rm p}^2/2}},
\end{equation}
for the pulse area to be $\pi$. As a consequence, $\delta t_{\rm p}$ is the quantity of interest to avoid overlap with the spectrum of the qubit emission. Ideally, one would want a narrow spectral pulse width (large $t_{\rm p}$) and a large separation (large $\delta$). But by increasing both of these quantities, $\Omega_0$ will correspondingly increase exponentially.

The second pumping scheme is a NARP pulse, proposed in~\cite{Wilbur2022}, which improves upon the original ARP technique~\cite{Loy1974,Malinovsky2001}. The original technique uses a chirped pulse to drive the system, which has a frequency which varies linearly with time. For a Gaussian envelope, the chirped pulse takes the form
\begin{equation}
    \Omega(t) = \Omega_0 e^{-t^2 /2 t_{\rm p}^2} e^{-i(\omega_{\rm{L}} + \alpha t/2)t},
    \label{ARP_pulseform}
\end{equation}
where $\Omega_0$ is chosen to achieve the desired pulse area, $t_{\rm p}$ is the pulse duration, $\omega_{\rm{L}}$ is the center frequency of the pulse, and $\alpha$ is the linear temporal chirp. 

The interaction Hamiltonian with the RWA is
\begin{equation}
    H_{\mathrm{S}}^{\rm NARP} = \hbar \Delta(t) \sigma^+ \sigma^- + \frac{\hbar\varepsilon(t)}{2} \left( \sigma^+ + \sigma^- \right),
\end{equation}
where we now have a time-dependent detuning between the pulse and TLS given by $\Delta(t) = \omega_0 - \omega_{\rm{L}} - \alpha t$, composed of a constant detuning, $\omega_0 - \omega_{\rm{L}}$, and $\alpha$ the rate of change of the detuning. Note we choose a positive chirp here to use the lower dressed state in the ARP method, avoiding phonon induced dephasing that occurs when going through the upper dressed state.

For ARP to work, the following conditions are required,
\begin{equation}
    |\alpha| t_{\rm p}^2 \gg 1,
\end{equation}
to ensure the change in frequency from the chirp is significant enough over the pulse duration of the Gaussian envelope, and
\begin{equation}
    |\alpha| t_{\rm p}^2 \ll \Omega_0^2 t_{\rm p}^2,
    \label{Cond2}
\end{equation}
which is the adiabatic condition for the changing dressed state energy splitting as $\Delta (t)$ varies. Notably, this means that NARP works for any pulse as long as Eq.~\eqref{Cond2} is satisfied.

The NARP scheme modifies the ARP pulse with an amplitude mask, $A(\omega)$, on top of the chirped pulse. The full mask (including the chirp) can be defined as $M(\omega) = A(\omega) e^{i \Phi(\omega)}$ and this gives the NARP pulse envelope to be
\begin{equation}
    \varepsilon'(t) = \mathcal{F}^{-1} [ \mathcal{F} [\varepsilon (t)] \cdot M(\omega) ],
\end{equation}
where $\mathcal{F}$ denotes the Fourier transform. The two masks take the form
\begin{equation}
    \begin{aligned}
        A(\omega) ={} & 1 - e^{- {\rm{ln}}(2) (\omega - \omega_{\rm{L}})^2 / \delta^2}, \\ 
        \Phi(\omega) ={} & \frac{\alpha'}{2} (\omega - \omega_{\rm{L}})^2,
    \end{aligned}
\end{equation}
where $2\delta$ is the FWHM of the spectral hole and $\alpha'$ is the spectral chirp (equivalent to the temporal chirp $\alpha$ but applied to the pulse in the frequency domain)
\begin{equation}
    \alpha' = \alpha \frac{t_{\rm p}^4}{1 + \alpha^2 t_{\rm p}^4}.
\end{equation}


The final driving scheme of interest is the SUPER scheme proposed by~\cite{Bracht2021} which uses two pulses far detuned from the qubit frequency to drive the qubit using the beating between the two pulses. The time dependent drive is
\begin{equation}
    \Omega(t) = \varepsilon'_1(t) e^{-i\omega_1 t} + \varepsilon'_2(t-\tau) e^{-i(\omega_2 t + \varphi)},
\end{equation}
with $\omega_1$ and $\omega_2$ the two frequencies of the input pulses and $\varphi$ a possible phase difference between them. The two pulse envelopes, $\varepsilon'_1$ and $\varepsilon'_2$, are both Gaussians with a possible time delay, $\tau$, between the two:
\begin{align}
    \varepsilon'_1(t) ={} & \Omega_1 e^{-t^2/2t_{\rm p,1}^2}, \\
    \varepsilon'_2(t) ={} & \Omega_2 e^{-(t-\tau)^2/2t_{\rm p,2}^2}.
\end{align}

Transforming the system Hamiltonian with this drive into the interaction picture at the first laser frequency, $\omega_1$, (with the RWA) yields
\begin{equation}
    \begin{aligned}
        H_{\mathrm{S}}^{\rm SUPER} = \hbar \Delta_1 \sigma^+ \sigma^- + \frac{\hbar \varepsilon'_1(t)}{2} \left(  \sigma^+ + \sigma^- \right) +{} & \\
        & \hspace{-5cm} \frac{\hbar \varepsilon'_2(t)}{2} \left(  e^{-i (\Delta_1 - \Delta_2)t} \sigma^+ + e^{i (\Delta_1 - \Delta_2)t} \sigma^- \right),
    \end{aligned}
\end{equation}
where $\Delta_i = \omega_0 - \omega_i$. This creates a dynamic where the qubit is pumped far off resonance by the first drive, while also being driven by the second laser with a fast oscillation at a frequency equal to the beating between the lasers. To ensure population inversion, the pulse detunings must be chosen such that
\begin{equation}
    \Delta_2 = \Delta_1 - \sqrt{\varepsilon_1^2(t=0) + \Delta_1^2}.
\end{equation}
This coordinates the beating between the lasers to be equal to the effective Rabi frequency at the peak of the first pulse to follow the required frequency switching in the detuned drive to complete inversion.

\subsection{Electron-Phonon Scattering Model}
\label{sec:PhononModel}

To take into account qubit dissipation and dephasing, we include a model for the QD-phonon reservoir coupling through a weak phonon coupling model~\cite{Gustin2020}. This model treats the exciton-phonon interaction, perturbatively, neglecting the non-Markovian parts of the scattering. The weak phonon model works quantitatively well at low temperatures \cite{Gustin2020}\footnote{Apart from a polaronic effect that can also easily by accounted for, caused by emission into the phonon sidebands, which is dicussed later.}, and for this work, we consider $T = 4$ K. This approach is especially effective here where the instantaneous pulse strengths can be very large causing a polaron transform approach (which includes phonons, non-perturbatively) to fail \cite{Mccutcheon2010,Mccutcheon2011}. 

The (weak) phonon master equation is
\begin{equation}
    \frac{d\rho}{dt} = \frac{-i}{\hbar} \left[ H_{\rm{S}}^x,\rho \right] + \mathbb{L}_{\rm{W}} \rho,
\end{equation}
where $\rho$ is the density matrix, $x$ denotes one of the three schemes, and 
\begin{equation}
    \begin{aligned}
        \mathbb{L}_{\rm{W}} \rho ={} & \int_0^{\infty} d\tau \Gamma_{\rm{W}} (\tau) \left( \tilde{N}(t-\tau,t) \rho N -{} \right. \\
        & \hspace{2cm} \left. N \tilde{N} (t-\tau,t) \rho \right) + \rm{H.c.}
    \end{aligned}
\end{equation}
is the phonon coupling term corresponding to exciton-phonon scattering, with $N = \sigma^+ \sigma^-$ and 
\begin{equation}
    \tilde{N} (t-\tau,t) = U^{\dagger} (t-\tau,t) N U(t-\tau,t),
\end{equation}
where $U(t,t_0)$ evolves the system from $t_0$ to $t$.

Making a Markov approximation, the coupling term is
\begin{equation}
    \mathbb{L}_{\rm{W}} \rho = \int_0^{\infty} d\tau \Gamma_{\rm{W}} (\tau) \left( \tilde{N}(-\tau) \rho N - N \tilde{N} (-\tau) \rho \right) + \rm{H.c.}
\end{equation}
with
\begin{equation}
    \tilde{N} (-\tau) = e^{-i H_{\rm{S}}^x(t) \tau/\hbar} N e^{i H_{\rm{S}}^x(t) \tau/\hbar},
\end{equation}
and
\begin{equation}
\begin{aligned}
    \Gamma_{\rm{W}} (\tau) ={} & \int_0^{\infty} d\omega J_P(\omega) \left[ {\rm{coth}} \left( \frac{\hbar \omega}{2 k_B T} \right) \cos{(\omega\tau)} -{} \right. \\
    & \hspace{2.5cm} \left. i \sin{(\omega \tau)} \right],
\end{aligned}
\end{equation}
where $T$ is the temperature and 
\begin{equation}
J_P(\omega) = \alpha_{\rm ph} \omega^3 {\rm{exp}} 
\left [- \frac{\omega^2}{2 \omega_b^2} \right ],
\end{equation}
is the phonon spectral function. The parameters $\alpha_{\rm ph}$ and $\omega_b$ are the exciton-phonon coupling strength and the cutoff frequency, respectively, which depend on the physical properties of the qubit. For an InGaAs/GaAs QD, such as those in~\cite{Quilter2015}, these parameters are $\alpha_{\rm ph} = 0.03 {\mathrm{ps}}^2$ and $\hbar \omega_b = 0.9 {\mathrm{meV}}$. This form of the spectral function represents bulk LA phonons in the continuum limit of the phonon spectral modes for a deformational potential in the QD lattice material~\cite{Nazir2016}. This spectral form has been well used to explain phonon effects in various QD experiments \cite{Ramsay2010,PhysRevB.86.241304,Wilbur2022,Dusanowski2022,Bracht2022,Bracht2023}.

Following the methods of~\cite{Gustin2020}, this electron-phonon coupling can be mapped onto a polaron shift of the qubit frequency $\Delta_P = \alpha_{\rm ph} \omega_b^3 \sqrt{\pi/2}$ and two phonon-mediated Lindblad operators, one equivalent to pure dephasing with rate $\gamma_{\rm{eff}}'(t)/2$ and operator $\sigma^+ \sigma^-$ and another {\it incoherent} excitation process which acts as phonon mediated excitation with rate $\Gamma^+(t)$ and operator $\sigma^+$. Importantly, both of these rates are a function of temperature. Specifically, these phonon-induced rates are given by
\begin{align}
    \gamma_{\rm{eff}}'(t) &= \pi \left( \frac{\varepsilon'(t)}{\varepsilon'_R(t)} \right)^2 J_P (\varepsilon'_R(t)) {\rm{coth}} \left( \frac{\hbar\varepsilon'_R(t)}{2 k_B T} \right), \label{phonon_pd} \\
    \Gamma^+(t) &= \frac{\pi}{4} \frac{|\varepsilon'(t)|}{\varepsilon'_R(t)} J_P (\varepsilon'_R(t)), \label{phonon_inv}
\end{align}
where $\varepsilon'_R(t) = \sqrt{\varepsilon'(t)^2 + \Delta^2}$ ($\Delta = \omega_0 - \omega_{\rm L}$). 
Note the absolute value in Eq.~\eqref{phonon_inv} is included to ensure a positive rate at all times as some of the driving schemes have negative (and real) instantaneous Rabi frequencies.
These phonon rates are time-dependent and their value depends on the strength and profile of the pump fields. 


By including the pump-induced phonon coupling, this introduces an inherent maximum to the indistinguishability that can be achieved. This maximum is due to the photon emission into the phonon sideband rather than the sharp zero phonon line and is determined by $\braket{B}^4$~\cite{Iles-Smith2017}, where $\braket{B}$ is the Franck-Condon factor:
\begin{equation}
    \braket{B} = \exp \left[ -0.5 \int_0^\infty d\omega \frac{J_P(\omega)}{\omega^2} \coth \left( \frac{\hbar \omega}{2 k_B T} \right) \right].
\end{equation}
For the parameter choices above, this gives $\braket{B} =0.96$ and a resulting cap on the indistinguishability of $0.849$. This can be improved through spectral filtering of the qubit emission but can result in a decrease of the efficiency of the SPS~\cite{Iles-Smith2017}. To get the clearest comparison between excitation schemes, we do not account for any specific filtering (which is typically used in experiments) and the indistinguishability limit dictated by $\braket{B}$, but highlight that any realization of these schemes will be limited by this effect.

\subsection{Simulation Model}
\label{sec:SimModel}

To simulate these systems and calculate their performance as a single photon source we use quantum trajectory theory~\cite{Dum1992,Tian1992,Dalibard1992}, a stochastic method which evolves the system ket vector with a non-Hermitian Hamiltonian augmented by quantum jumps. Using quantum trajectories has some significant numerical advantages when calculating the single photon source figures of merit as we now describe.

Conventionally, open system quantum optics simulations are done using a master equation approach. For a pulsed system, calculating $g^{(2)}(0)$ and $\mathcal{I}$ with the master equation requires simulating the full two time correlation functions $g^{(1)}(t,t')$ and $g^{(2)}(t,t')$. These simulations scale as $D^2 \cdot T^2$ where $D$ is the Hilbert space dimension and $T$ is the required number of time steps for the system to complete relaxation after the pulse.

In the quantum trajectory picture, we use the natural link between the trajectories and experimental realizations of the system to simulate Hanbury Brown and Twiss (HBT) and Hong-Ou-Mandel (HOM) interferometers~\cite{Crowder2024}. In contrast, the HBT simulations scale as $N \cdot D \cdot T$ and the HOM simulations scale as $N \cdot D^2 \cdot T$ where $N$ is the number of trajectories required to converge to the ensemble average. Typically this is on the order of a few thousand trajectories, while the required time steps can be on the order of thousands to hundreds of thousands depending on the time dynamics that must be resolved. Furthermore, each simulation is independent and so parallel computing can be used to further reduce the computation time by up to a factor of $N$ (depending on the number of computing nodes available).

\section{Numerical Results}
\label{sec:Results}

To compare the performance of the three driving schemes, we simulate the common QD system driven by each scheme for single photon emission. We quantify the SPS performance of the system through three figures of merit; the efficiency, $\eta$, of single photon emission, the coherence, $g^{(2)}(0)$, of the emitted photons, and the indistinguishability, $\mathcal{I}$, of the emitted photons.

\begin{figure*}
    \centering
    \includegraphics[width=1.5\columnwidth]{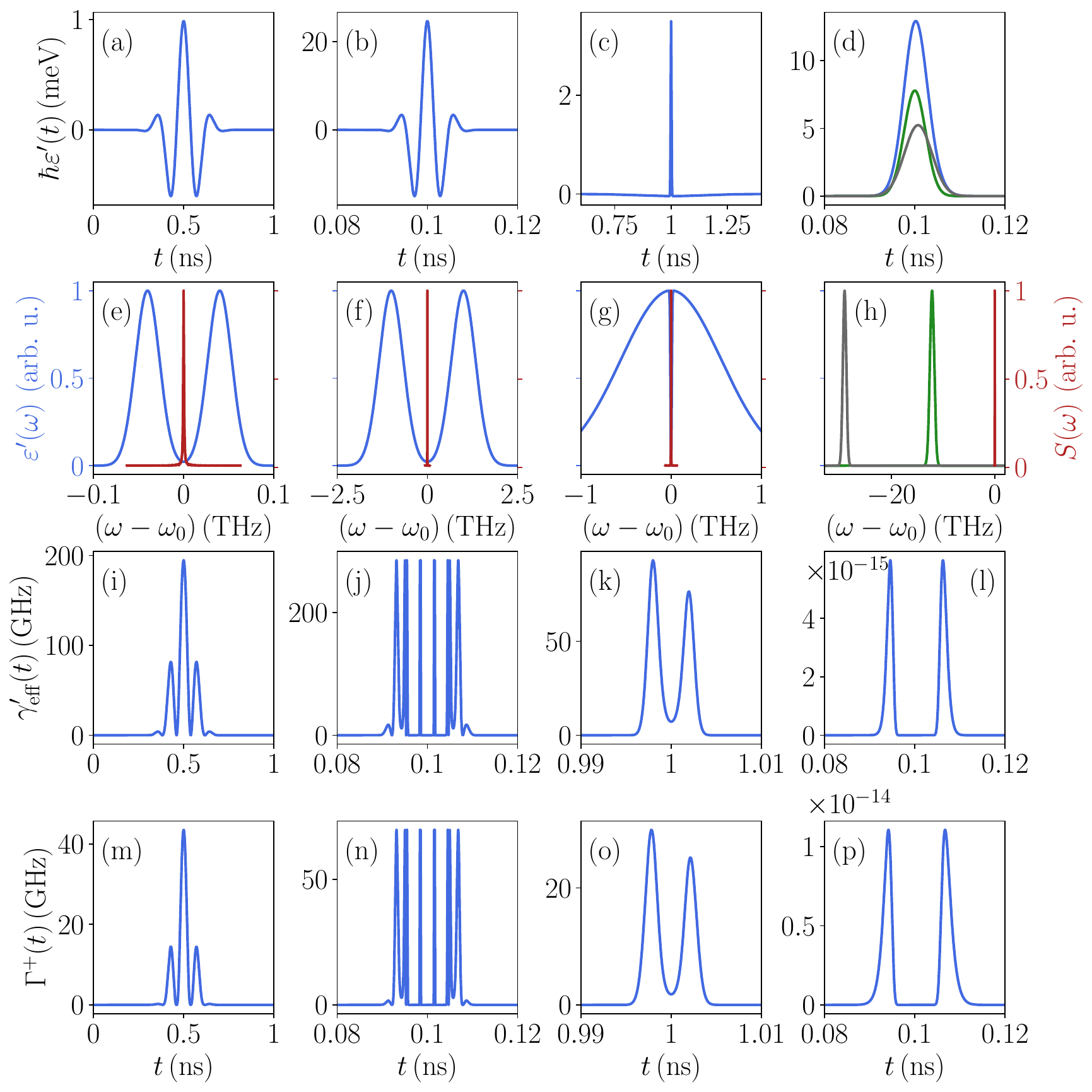}
    \vspace{-0.5cm}
    \caption{The laser profiles and resulting phonon rates for the four driving schemes in each column, left to right: long dichromatic pulse, short dichromatic pulse, NARP, and SUPER. By row, (a)-(d) the time domain profile of the pulse, (e)-(h) the spectral profile (in blue) with the qubit emission spectrum (in red), (i)-(l) the phonon-induced dephasing rate, and (m)-(p) the phonon-induced excitation rate.
    }
    \label{DriveProfiles}
\end{figure*}

In this section, we use $\gamma/2\pi = 1$ GHz ($\hbar \gamma \approx 4.1~\mu$eV) similar to the QD emission rate in~\cite{Reimer2016}, but note that all of the results are sensitive to the pulse parameters relative to $\gamma$. For example, the rate of two photon emission, a detriment to perfect single photon generation, is dependent on the effective pulse width of the pumping scheme relative to $\gamma$. Further, the full width half maximum of the qubit emission spectrum (shown in red in Fig.~\ref{DriveProfiles}) is $\gamma$ and thus the overlap between the qubit emission and the laser spectrum also depends on this value. An important feature of each of these off-resonant pumping schemes is that they avoid spectral overlap with the qubit emission spectrum. To quantify this, we define an overlap measure,
\begin{equation}
    \mathcal{O} = \frac{\int |\varepsilon'_{\rm{norm}}(\omega)|^2 S(\omega) d\omega}{\int S(\omega) d\omega},
\end{equation}
which describes the overlap between the spectral function of the (normalized) input driving
spectrum, $|\varepsilon'_{\rm{norm}}(\omega)|^2$, and the qubit emission spectrum, $S(\omega)$.

The parameters for each pulse that we choose are intended to most fairly compare the schemes using parameters proposed in the original papers, 
with similar pulse durations for each scheme. The parameters chosen are listed in Table~\ref{Driving_Params}. Two parameters sets are shown for the dichromatic pumping scheme because the first set of parameters taken from the original paper~\cite{He2019} proposes a much larger pulse width than in either the NARP or SUPER schemes. 

\begin{table*}[]
\begin{tabular}{|c|c|c|c|c|c|c|}
\hline
\multicolumn{1}{|c|}{Scheme} & $\hbar\Omega_0$  (meV) & $t_{\rm p}$ (ps) & $\varphi$ & $\hbar\Delta(t=0)$ (meV) & $ \hbar\delta$ ($\mu$eV)     & Other Parameters \\ \hline
Long Dichromatic & $0.314$ & $74.96$ & $0$ & N/A & $26.3$ & N/A \\ \hline
Short Dichromatic & $7.842$ & $3.00$ & $0$ & N/A & $658.2$ & N/A \\ \hline
NARP & $3.547$ & $1.80$ & N/A & $0$ & $5.3$ & $\alpha = -1.111 \, \mathrm{ps}^{-2}$ \\ \hline
SUPER & \begin{tabular}[c]{@{}l@{}}$\Omega_1 : 7.785$\\ $\Omega_2 : 5.235$\end{tabular} & \begin{tabular}[c]{@{}l@{}}$t_{\rm p,1} : 2.40$\\ $t_{\rm p,2}: 3.04$\end{tabular} & $0$       & \begin{tabular}[c]{@{}l@{}}$\Delta_1 : 8.00$\\ $\Delta_2 : 19.163$\end{tabular} & N/A          & $\tau = -0.73$ ps     \\ \hline
\end{tabular}
\caption{Pulse profile parameters for the four driving setups.}
\label{Driving_Params}
\end{table*}

We first compare the drive profiles in time and frequency for the four driving setups in Fig.~\ref{DriveProfiles}(a)-(h). In the time domain (a - d), the SUPER scheme (d) and the short dichromatic pulse (b) both have very high instantaneous Rabi frequencies during the pulse. This yields very fast time dynamics in the qubit evolution, while the NARP (c) and long dichromatic pulse (a) have instantaneous Rabi frequencies about an order of magnitude lower. In the frequency domain (e - h), all of the pulses avoid significant overlap with the qubit emission spectrum, however some achieve this better than others. The SUPER scheme (h) is best for reducing spectral overlap, as the detuning of the two composite pulses results in no overlap between the spectra, $\mathcal{O} = 0$. The dichromatic scheme pulse spectra in (e) and (f) may look identical, but the shorter pulse width in (b) causes the two pulses in (f) to be an order of magnitude wider than in (e). This results in an overlap of $\mathcal{O} = 0.0049$ for the longer pulse in (e) and an overlap of $\mathcal{O} = 0.0005$ for the shorter pulse in (f). This is non-zero because the tails of the detuned Gaussians still cover the frequency of the qubit. Lastly the NARP pulse (g) has a very broad spectrum so it looks like it overlaps with the qubit emission, but the sharp magnitude filter, $A(\omega)$, cuts out the portion of the spectra about the qubit emission and has an overlap of $\mathcal{O} = 0.0263$.

For the qubit model considered in this work, there are three mechanisms which lead to imperfect single photon source emission. The first is multiphoton emission that occurs during the excitation process, i.e., the qubit emits before the laser has finished driving it to the excited state. This is primarily mitigated through a shorter pulse duration and is why we included a shorter version of the dichromatic pulse to have comparable rates of multiphoton emission for the three schemes. 

The second and third are phonon-induced pure dephasing and excitation, with rates given in Eqs.~\eqref{phonon_pd} and~\eqref{phonon_inv}, which (with respect to the driving pulse) are determined by the strength and detuning of the pulse; they are calculated in the time domain for each of the four pulses in Figs.~\ref{DriveProfiles}(i)-(p).  High instantaneous rates are seen for the short dichromatic pulse, but they are interspersed with very low rates. This happens because the high Rabi frequency surpasses the cutoff frequency of the phonon spectrum and where this is true the $\rm{exp}[-(\varepsilon_R')^2/2\omega_b^2]$ term suppresses the influence of phonons. This also occurs for the SUPER scheme, but the extremely detuned pulses cause the phonon rates to be essentially zero ($\propto 10^{-14}$ GHz) and phonons have very little effect on this driving scheme. The longer dichromatic pulse also has phonon rates $>100$ GHz and because of the longer interaction time this can completely spoil the excited state preparation as we will show. By using a positive chirp, the higher side of the phonon rates for the NARP scheme are before excitation has begun to occur in the excitation scheme which improves the SPS performance. Further, as it passes through the peak of the pulse, the phonon rates drop off significantly, avoiding dissipation during the most sensitive part of the excitation process.

\begin{figure*}
    \centering
    \includegraphics[width=1.8\columnwidth]{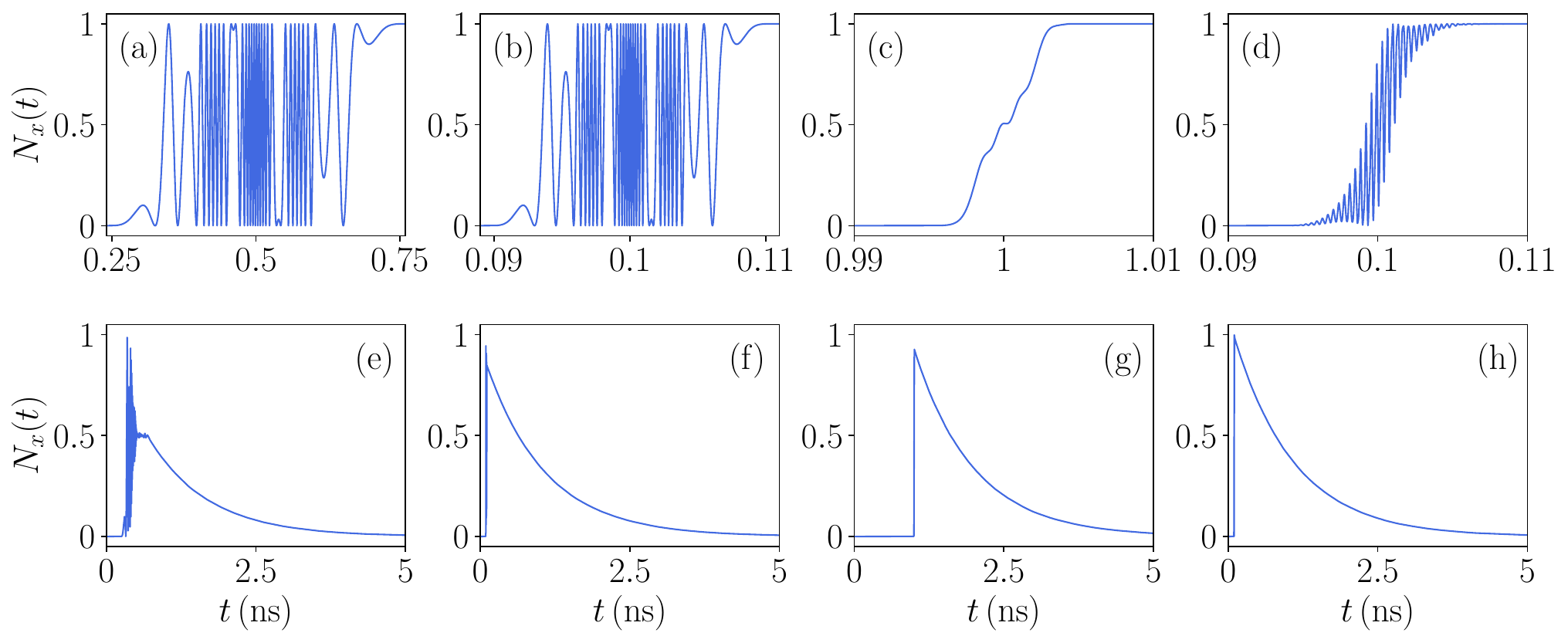}
    \vspace{-0.5cm}
    \caption{The population dynamics ($N_x(t) = \braket{\sigma^+ \sigma^- (t)}$) without dissipation in the first row, and with waveguide output and phonon coupling in the second row. Each column is a particular driving scheme: (a)-(b) long dichromatic pulse, (c)-(d) short dichromatic pulse, (e)-(f) NARP, and (g)-(h) SUPER.}
    \label{PopDynam}
\end{figure*}

Next in Fig.~\ref{PopDynam}, we show how the population for each scheme evolves without dissipation and with phonon coupling. In the first row, without dissipation, all schemes successfully reach $>99.96\%$ population inversion. Notably, only the dichromatic scheme does this to exactly $100\%$ because it analytically satisfies the area theorem. Both the NARP and SUPER schemes are threshold approaches and only achieve near perfect inversion but do not exactly reach the excited state. In the second row, the population dynamics including both decay into the waveguide and phonon coupling are shown for each scheme. The SPS figures of merit for each pumping scheme are given in Table~\ref{SPS_FoM}. As expected from the analysis of the phonon rates, the long dichromatic pulse loses coherence due to the high rate of pure dephasing and results in a maximum population of 0.5 before the pulse has completed. This severely limits its usefulness as a single photon source as the efficiency of the source is capped at 0.5, with associated hits to its coherence and indistinguishability. In contrast, the short dichromatic pulse avoids this large phonon influence and the dynamics show a promising single photon source. It does take a hit to its efficiency ($\eta = 0.8519$) but has very good coherence ($g^{(2)}(0) = 0.0074$) and indistinguishability ($\mathcal{I} = 0.9889$).

Both the NARP and SUPER schemes result in near perfect single photons with $g^{(2)}(0) = 0.0036$ and $\mathcal{I} = 0.9929$ for the NARP scheme and $g^{(2)}(0) = 0.0012$ and $\mathcal{I} = 0.9987$ for the SUPER scheme. The main differentiator between the two is their efficiency which are both high but is better for the SUPER scheme ($\eta = 0.9890$) than the NARP scheme ($\eta = 0.9302$). For near perfect single photons, it is surprising that the efficiency takes such a hit in the NARP case. The phonon-induced excitation events are the culprit for this behaviour (and also for the outsize hit to the efficiency in the short dichromatic pulse as well). Before the driving pulse begins to coherently excite the dot, there is a peak in the phonon-induced excitation rate. If such an excitation event occurs, the qubit begins in the excited state as the driving pulse arrives and the pulse acts to de-excite the qubit from the excited state to the ground state, resulting in no emission. This gives a pulse for which no photon is emitted (decreasing the efficiency) but does not appear in the HBT or HOM interferometric histograms which are used to calculate coherence and indistinguishability.

\begin{table}[]
\begin{tabular}{|l|l|l|l|}
\hline
\multicolumn{1}{|c|}{Scheme} & \multicolumn{1}{c|}{$\eta$} & \multicolumn{1}{c|}{$g^{(2)}(0)$} & \multicolumn{1}{c|}{$\mathcal{I}$} \\ \hline
Long Dichromatic & 0.4948 & 0.4619 & 0.5850 \\ \hline
Short Dichromatic & 0.8519 & 0.0074 & 0.9889 \\ \hline
NARP & 0.9302 & 0.0036 & 0.9929 \\ \hline
SUPER & 0.9890 & 0.0012 & 0.9987 \\ \hline
\end{tabular}
\caption{SPS figures of merit for the four population dynamics with phonon coupling ($\alpha_{\rm ph} = 0.03~ {\mathrm{ps}}^2$ and $\hbar \omega_b = 0.9~ {\mathrm{meV}}$) in the second row of Fig.~\ref{PopDynam}.
}
\label{SPS_FoM}
\end{table}

\begin{figure}
    \centering
    \includegraphics[width=0.99\columnwidth]{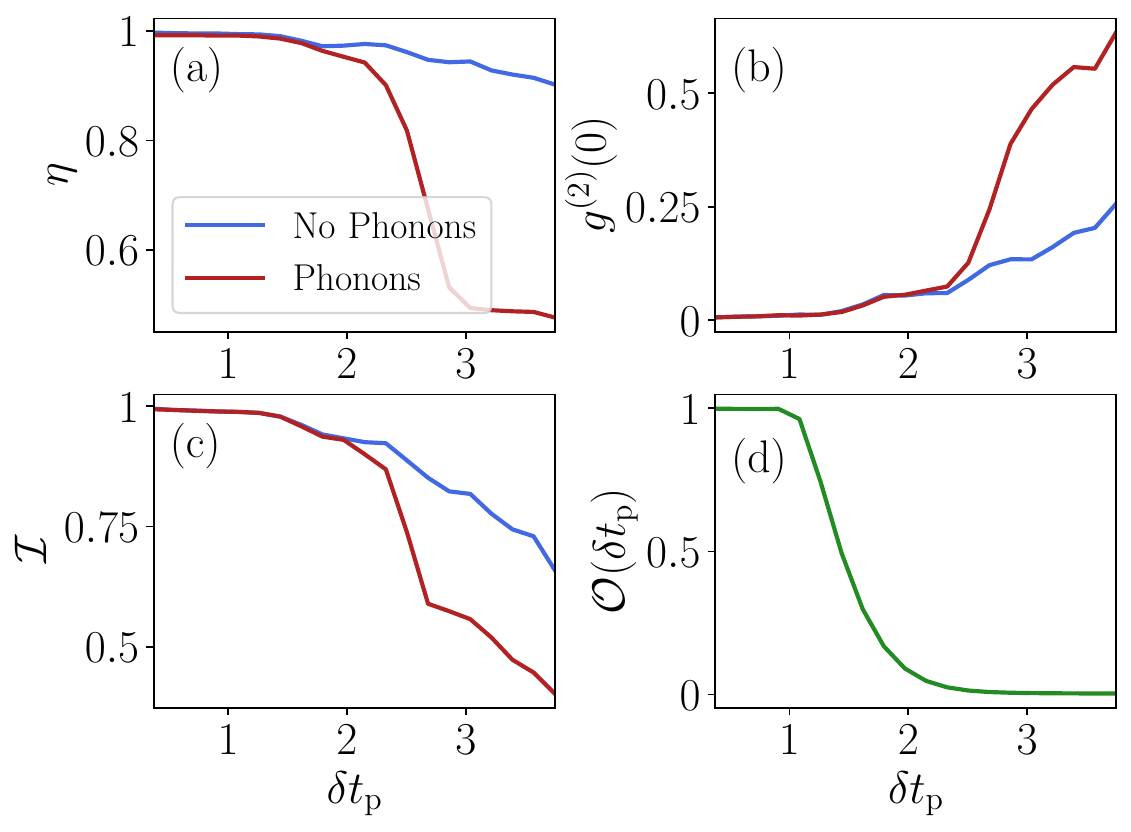}
    \vspace{-0.5cm}
    \caption{(a)-(c) SPS performance (through the (a) efficiency, (b) coherence, and (c) indistinguishability) as a function of the spectral separation of the two pulses for the long dichromatic pulse. The overlap function for each separation is in (d). Note that the overlap results in (d) are only dependent on the pulse form.}
    \label{GapVary_Dichrom_Long}
\end{figure}

\begin{figure}
    \centering
    \includegraphics[width=0.99\columnwidth]{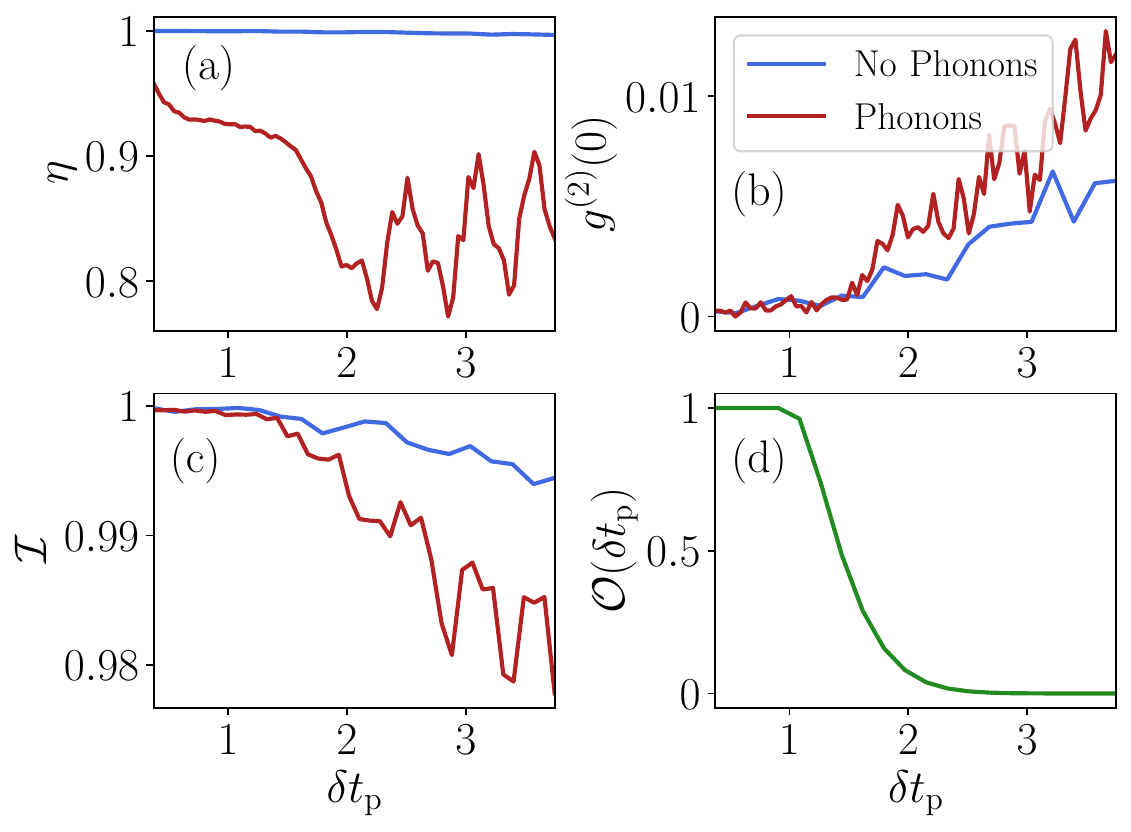}
    \vspace{-0.5cm}
    \caption{(a)-(c) SPS performance (through the (a) efficiency, (b) coherence, and (c) indistinguishability) as a function of the spectral separation of the two pulses for the short dichromatic pulse. The overlap function for each separation is in (d). Each data point in (a)-(c) is an average of 50,000 stochastic trajectories with very low event rates which leads to the noise in the data.
    }
    \label{GapVary_Dichrom_Short}
\end{figure}

It is also interesting to compare the SPS performance of the dichromatic and NARP schemes as their spectral gap increases (which decreases the overlap function). For the dichromatic pulses the spectral gap is determined by the value of $\delta t_{\rm p}$, and for the NARP pulse it is determined by $\delta$. The SPS performance vs. the spectral gap size is shown for the long dichromatic pulse in Fig.~\ref{GapVary_Dichrom_Long} and for the short dichromatic pulse in Fig.~\ref{GapVary_Dichrom_Short}. Similar changes in the performance are seen in both cases, with a larger hit to performance being seen in the case of the longer dichromatic pulses. All three figures of merit; efficiency, coherence, and indistinguishability decrease as the detuning between the two pulses increases both with and without phonon coupling. As the detuning increases, the duration of the total pulse duration also increases, in turn causing an increase in multiphoton emission. This is the cause of the performance decrease without phonons. Additionally, as the detuning increases, the amplitude of the laser pulse also increases, leading to larger rates of phonon dephasing and excitation which cause the additional impact in performance.

\begin{figure}
    \centering
    \includegraphics[width=0.99\columnwidth]{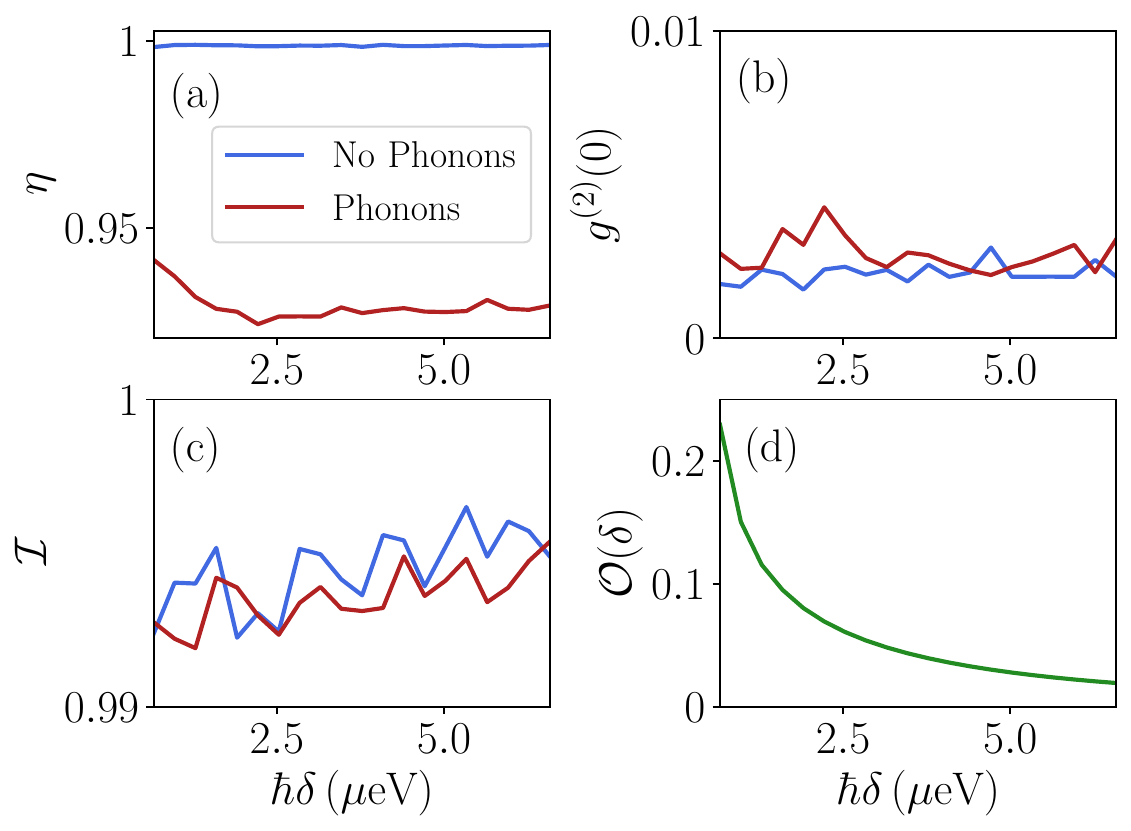}
    \vspace{-0.5cm}
    \caption{(a)-(c) SPS performance (through the (a) efficiency, (b) coherence, and (c) indistinguishability) as a function of the spectral filter width in NARP. The overlap function for each separation is in (d). Each data point in (a)-(c) is an average of 50,000 stochastic trajectories with very low event rates which leads to the noise in the data.}
    \label{GapVary_NARP}
\end{figure}

Figure~\ref{GapVary_NARP} shows the SPS performance as the amplitude mask of the NARP pulse widens. The most important feature here is that as the mask widens, there is no significant impact on the performance in either of the three figures of merit. Aside from the hit to efficiency from phonon-induced excitation, one can use an amplitude mask as wide as desired to avoid spectral overlap with no resulting loss in performance. This can be experimentally useful as the pulse scheme will continue to work within a large margin of error on the laser parameters. Also, by allowing for very wide filters, one can avoid driving any additional resonances that may be present in the qubit.

\begin{figure*}
    \centering
    \includegraphics[width=1.8\columnwidth]{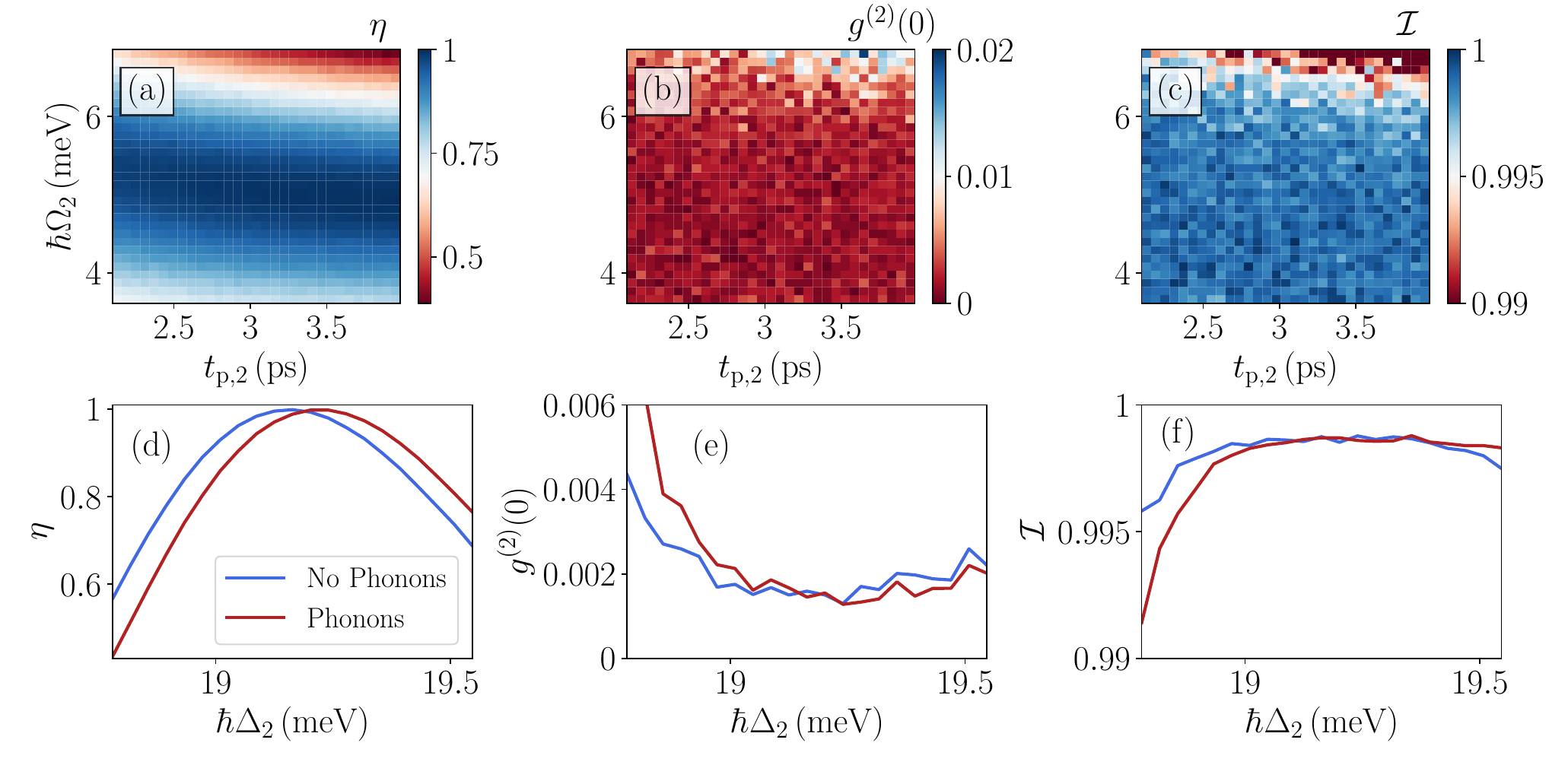}
    \vspace{-0.3cm}
    \caption{(a)-(c) SPS performance as a function of the amplitude ($\Omega_2$) and the pulse width ($t_{\rm p,2}$) of the second laser in the SUPER scheme. Each parameter is varied to a maximum of 30\%. (d)-(f) SPS performance as a function of the second laser detuning.
    }
    \label{SUPER_CombinedScans}
\end{figure*}

It is important to note that the laser parameters for the SUPER scheme are not analytically derived. Rather the parameters of the first laser are fixed and then a scan of the second laser parameters are used to find parameter sets which satisfy inversion. Here, we compare the SPS performance for the parameter space around the SUPER laser parameters for the second pulse used in the previous analysis. In Fig.~\ref{SUPER_CombinedScans}(a)-(c), we scan the parameter space of the second laser's envelope by varying the amplitude and pulse width up to 30\% (so the shown parameter space is $[0.7\Omega_2,1.3\Omega_2]$ and $[0.7t_{\rm p,2},1.3t_{\rm p,2}]$). In (b) and (c), the coherence and indistinguishability show no underlying pattern as either parameter changes showing no significant impact on the performance. However, in the efficiency, there is a hit as the second laser varies its envelope. There is a small change as $t_{\rm p,2}$ increases or decreases, with the efficiency only dropping to $\eta = 0.973$ after 30\% decrease in $t_{\rm p,2}$. A much larger effect appears as $\Omega_2$ changes with a 5\% increase in amplitude results in an efficiency of $\eta = 0.96$ and a 20\% increase results in an efficiency of $\eta = 0.718$. 

In Fig.~\ref{SUPER_CombinedScans}(d)-(f), the detuning of the second pulse, $\Delta_2$, is varied up to 2\%, a shift of 0.38 meV. When the detuning is shifted, the efficiency quickly falls off to $<$0.9 after a shift of only 0.19 meV, while the coherence and indistinguishability still remain largely unaffected. This is because if the detuning is incorrect, the fast oscillations will no longer result in the beating driving the TLS to the excited state, but rather it is out of rhythm with the Rabi oscillations of the first laser and inversion fails. This however, does not impact the coherence and indistinguishability as no additional multiphoton or dephasing events are occurring to spoil these figures of merit.

\vspace{0.2cm}
\section{Conclusions}
\label{sec:Conc}

In summary, we have compared the SPS performance through the efficiency, coherence, and indistinguishability of the emitted photons from three different off resonant excitation schemes driving a realistic waveguide QED qubit model. While the dichromatic scheme can achieve reasonable success as a single photon source, its performance falls as the spectral spacing between the pulses widens (reducing the spectral overlap between qubit and laser) and phonon dissipation begins to negatively impact the SPS performance. 

Both the NARP and SUPER schemes achieve near perfect coherence and indistinguishability. While both achieve excellent efficiency, the SUPER scheme is slightly ahead of the NARP scheme in this figure of merit. The NARP scheme is also quite robust against any variance in the driving parameters and can have the spectral filter broadened to any needed width with little to no impact of the SPS performance. The SUPER scheme is quite sensitive to the parameters of its constituent lasers, with a 2\% change in the detuning of the second laser resulting in a hit of over 50\% to the efficiency and changes of up to 5\% in the amplitude and width causing a 3\% decrease in efficiency.

These pulse schemes allow efficient SPS preparation without the need for polarization filtering allowing for the complete single photon pulse to be used rather than a single polarization. Further work can be done to optimize these schemes within the free parameters of each scheme, or beyond that, inverse design techniques can be used that could create a pulse with better and more robust SPS performance while avoiding overlap with the qubit emission spectrum.

\vspace{0.5cm}

\acknowledgements
This work was supported by the Natural Sciences and Engineering Research Council of Canada (NSERC), the Canadian Foundation for Innovation (CFI), the National Research Council of Canada (NRC), Queen's University, Canada, and the University of Ottawa.

\bibliography{refs}

\end{document}